\documentclass[aps,preprint,nofootinbib]{revtex4-1}%
\usepackage{amsfonts}
\usepackage{amsmath}
\usepackage{amssymb}
\usepackage{graphicx}%
\begin{document}
\title{Trimaximal-Cabibbo neutrino mixing: A parametrization in terms of deviations
from tri-bimaximal mixing}
\author{Bo Hu}
\email{bohu@ncu.edu.cn}
\affiliation{Department of Physics, Nanchang University, Nanchang 330031, China}

\begin{abstract}
In this paper we study a parametrized description of neutrino mixing from a
phenomenological point of view. We concentrate on the parametrization in terms
of higher order corrections to the leading order mixing matrix. A method to
describe subleading contributions and its applications to tri-bimaximal mixing
are discussed. We show that mixing matrices similar to tri-bimaximal-Cabibbo
mixing can be obtained by straightforward choices of parameters. To achieve
better agreement with the experimental data without increasing the number of
free parameters, we impose a simple phenomenological relation from which a
trimaximal-like mixing matrix, parametrized by $U_{e3}=\sin \theta
_{13}e^{-i\varphi}$, can be derived straightforwardly without imposing
additional requirements. It can describe the current global fit to
three-neutrino mixing with good accuracy. Its theoretical explanation and
phenomenological applications are discussed briefly.

\end{abstract}
\maketitle

\section{Introduction}

Neutrino mixing is one of the most extensively studied topics in neutrino
physics. The mixing pattern observed in neutrino oscillation experiments
provides clear evidence, implying a non-trivial but perhaps simple flavor
structure of the lepton sector. Many interesting mixing patterns are proposed
to describe the mixing data, including tri-bimaximal mixing (TBM) \cite{TBM},
bimaximal mixing \cite{BM}, golden-ratio mixings \cite{GR}, democratic mixing
\cite{DMM} and hexagonal mixing \cite{HM}, etc. Some of these often appear as
exact mixing matrices\footnote{By "exact mixing matrix", we mean the mixing
matrix that does not depend on other parameters, including\ lepton masses. The
corresponding effective neutrino mass matrix is sometimes called the "form
diagonalizable matrix" \cite{fm1, fm2}.} in models where neutrino mixing is
determined by underlying discrete flavor symmetries (see, e.g. \cite{dsr, dm}).

This paradigm worked quite\ well\ before the recent discovery of a relatively
large $\theta_{13}$ (for recent global fits, see \cite{Tortola, Fogli,
Gonzalez}), which signals a deviation from those exact mixing patterns. For
example,\ a recent global fit by M.~C. Gonzalez-Garcia \emph{et al}.
\cite{Gonzalez} gives
\begin{align}
\sin^{2}\theta_{12} &  =0.30\pm0.013,\nonumber \\
\sin^{2}\theta_{23} &  =0.41_{-0.025}^{+0.037}\oplus0.59_{-0.022}%
^{+0.021},\label{gfit}\\
\sin^{2}\theta_{13} &  =0.023\pm0.0023,\nonumber
\end{align}
where two $1\sigma$ ranges for $\sin^{2}\theta_{23}$ are given because the
present data cannot resolve the $\theta_{23}$ octant degeneracy
\cite{Gonzalez}. In the global fit given above, $\theta_{ij}$ are the mixing
angles defined in the standard parametrization. From Eq.~(\ref{gfit}), the
squared mixing matrix elements $|(U_{\nu})_{ij}|^{2}$ can be calculated and
given collectively in a matrix as%

\begin{equation}
\left \Vert |(U_{\nu})_{ij}|^{2}\right \Vert =\left(
\begin{array}
[c]{ccc}%
0.684_{-0.013}^{+0.013} & 0.293_{-0.013}^{+0.013} & 0.023_{-0.0023}%
^{+0.0023}\\
\cdots & \cdots & 0.401_{-0.024}^{+0.036}\oplus0.576_{-0.022}^{+0.021}\\
\cdots & \cdots & 0.576_{-0.036}^{+0.024}\oplus0.401_{-0.021}^{+0.022}%
\end{array}
\right)  \label{uij2}%
\end{equation}
in which $|(U_{\nu})_{ij}|^{2}$ with $i\neq e$ and $j\neq3$ are\ omitted
because they are affected by the Dirac CP phase whose experimental\ value has
a relatively large error. In our discussion they are determined by other
parameters and, hence, the CP phase can be extracted by standard formalism.

Although most exact mixing patterns including those mentioned above are not in
precise agreement with the experimental data, improvement can be made by
introducing small but non-negligible subleading contributions, which can be
induced by radiative corrections, charged lepton corrections, etc. (see, e.g.,
\cite{cc}).\ With this in mind, the parametrization of neutrino mixing by
deviations from leading order (LO) exact mixing patterns has been studied
extensively. For some relevant works, see \cite{King2, TMM, TMP, ppc} and
other references given in this paper.

In the existing literature, a common choice of parameters is the deviation of
mixing angles and the Dirac CP phase.\footnote{Majorana phases can be ignored
since they do not affect neutrino oscillations \cite{Xing2}.} In this paper,
we use a different method where the mixing matrix is given by the product of a
matrix describing deviations and the LO mixing matrix. In some cases it is
simpler than the method dealing with mixing angles, and the physical relevance
is more transparent. Although this method can be used for any LO\ mixing, we
shall concentrate on tri-bimaximal mixing in this paper. We show that mixing
matrices similar to tri-bimaximal-Cabibbo (TBC) mixing \cite{King2} can be
obtained by this method with straightforward choices of parameters.

However, these TBC-like mixing patterns agree with the data only marginally.
To achieve better agreement, sizable corrections to $\theta_{12}$ and
$\theta_{23}$ must be taken into account. To do that, one can introduce more
parameters. Nevertheless, parametrizations with fewer parameters can lead to
simplified descriptions of neutrino phenomenologies and may provide clues to
underlying physics. Therefore, it is also worthwhile to look for and study
simple descriptions of neutrino mixing suggested by the experimental data. In
this respect, phenomenological or empirical relations are very useful, e.g.,
the relation between $\theta_{13}$ and the Cabibbo angle $\theta_{C}$:
$\theta_{13}\approx \theta_{C}/\sqrt{2}$ \cite{King2}, the quark-lepton
complementarity \cite{Sim2, Sim3}, self-complementarity \cite{Ma1, Ma2},
bi-large mixing \cite{bl}, bi-pair mixing \cite{bp}, etc. Besides their
theoretical implications, they may also give rise to economical but rather
accurate descriptions of neutrino mixing and phenomenologies.

In this paper, we introduce a relation, i.e., $|\left(  U_{\nu}\right)
_{e1}|^{2}-|\left(  U_{\nu}\right)  _{e3}|^{2}=2/3$, which agrees with the
data quite well, as can be seen from Eq.~(\ref{uij2}). We show that this
relation leads to a parametrization that can describe the current global fit
with good accuracy. It is derived from the result obtained in Sec.~II and
parametrized by $U_{e3}=\sin \theta_{13}e^{-i\varphi}$. Hence it can be
regarded as an improved TBC mixing. Moreover, it is also a trimaximal-like
mixing with the $\mathrm{TM}_{2}$ trimaximal condition \cite{TMM} being
perturbed by a small correction. Hence, we refer to it as trimaximal-Cabibbo mixing.

In summary, the paper is organized as follows. In Sec.~II, we introduce a
method to describe small deviations from LO mixing matrices. Several cases are
discussed in detail. In Secs.~III and IV, the results obtained in Sec.~II are
used to derive TBC-like mixing and trimaximal-Cabibbo mixing. In particular,
the latter and its derivation are discussed in greater detail in Sec.~IV. In
Sec. V, we summarize and discuss briefly the theoretical explanation and
phenomenological applications of trimaximal-Cabibbo mixing.

\section{Deviations from leading order mixing matrix}

In this section we discuss a method to describe deviations from LO mixing
matrices. Note that the lepton mixing matrix $U_{\nu}$ can always be written
as a product of two unitary matrices, i.e.,%
\begin{equation}
U_{\nu}=U_{\nu}^{0}T\label{mm0}%
\end{equation}
When $T=\mathbb{I}$, the identity matrix, one has $U_{\nu}=U_{\nu}^{0}$.
Hence, we use $U_{\nu}^{0}$ to denote the LO mixing matrix and the matrix $T$
to describe the deviations of $U_{\nu}$ from $U_{\nu}^{0}$. Just for
convenience, in the following the matrix $T$ will be referred to as a
perturbation matrix, although it is appropriate only when the deviations are
very small. The formalism developed in the following can also be used for the
case where the matrix $T$ multiplies $U_{\nu}^{0}$ from the left, i.e.,
$U_{\nu}=TU_{\nu}^{0}$, which will be discussed briefly at the end of this section.

In Eq.~(\ref{mm0}), the mixing matrix $U_{\nu}$ and the LO mixing matrix
$U_{\nu}^{0}$ can be written as%
\begin{equation}
U_{\nu}=(U_{1},U_{2},U_{3}),\;U_{\nu}^{0}=(K_{1},K_{2},K_{3}).\label{mm1}%
\end{equation}
where $U_{i}$ and $K_{i}$ for $i=1$, $2$, or $3$ are column vectors of
$U_{\nu}$ and $U_{\nu}^{0}$, i.e.%
\begin{equation}
U_{i}=\left(
\begin{array}
[c]{c}%
\left(  U_{\nu}\right)  _{ei}\\
\left(  U_{\nu}\right)  _{\mu i}\\
\left(  U_{\nu}\right)  _{\tau i}%
\end{array}
\right)  ,\quad K_{i}=\left(
\begin{array}
[c]{c}%
\left(  U_{\nu}^{0}\right)  _{ei}\\
\left(  U_{\nu}^{0}\right)  _{\mu i}\\
\left(  U_{\nu}^{0}\right)  _{\tau i}%
\end{array}
\right)  .\label{UK}%
\end{equation}
The perturbation matrix $T$ in Eq.~(\ref{mm0}) is a unitary matrix that can be
parametrized by three angles denoted by $\varsigma_{ij}$ and six phases. As
usual, angle $\varsigma_{ij}$ corresponds to the rotation angle in the $(i,j)$ plane.

We consider first the simplest case where $T$ has two vanishing angles. For
example, when $\varsigma_{12}=\varsigma_{23}=0$ and $\varsigma_{13}\neq0$, $T$
can be written as%
\begin{equation}
T=e^{i\omega}\left(
\begin{array}
[c]{ccc}%
\cos \varsigma_{13} & 0 & \sin \varsigma_{13}e^{-i\alpha}\\
0 & 1 & 0\\
-\sin \varsigma_{13}e^{i\alpha} & 0 & \cos \varsigma_{13}%
\end{array}
\right)  P_{\omega}\label{t13r}%
\end{equation}
where $P_{\omega}=\mathrm{diag}\{e^{i\omega_{1}},1,e^{i\omega_{2}}\}$. Since
$P_{\omega}$ and $e^{i\omega}$ can be adsorbed by Majorana phases or charged
leptons, we will ignore them in the following discussions. Thus, from
Eqs.~(\ref{mm0}), (\ref{mm1}), and (\ref{t13r}) one has
\begin{equation}
U_{1}=(K_{1}-x^{\ast}K_{3})/f_{1},\quad U_{2}=K_{2},\quad U_{3}=(K_{3}%
+xK_{1})/f_{3}\label{unu13}%
\end{equation}
where%
\[
f_{1}=f_{3}=\sqrt{1+|x|^{2}},\;x=\tan \varsigma_{13}e^{-i\alpha}%
\]
As we can see, instead of angles and phases, one can also use $x$ to
parametrize $U_{\nu}$.

The above procedure can be applied consecutively and iteratively. Below we
consider two cases that are relevant to later discussions. In the first case
the perturbation matrix $T$ is given by a rotation in $(1,3)$ plane followed
by a rotation in $(1,2)$ plane. By applying Eq.~(\ref{unu13}) twice on these
rotations, it is straightforward to find that%
\begin{align}
U_{1} &  =\left(  K_{1}-y^{\ast}f_{3}K_{2}-x^{\ast}K_{3}\right)
/f_{1},\nonumber \\
U_{2} &  =\left(  f_{3}K_{2}+yK_{1}-x^{\ast}yK_{3}\right)  /f_{2}%
,\label{unu1312}\\
U_{3} &  =\left(  K_{3}+xK_{1}\right)  /f_{3},\nonumber
\end{align}
where%
\begin{equation}
f_{1}=f_{2}=\sqrt{(1+|x|^{2})(1+|y|^{2})},\quad f_{3}=\sqrt{1+|x|^{2}%
}.\label{unu1312n}%
\end{equation}
Equations (\ref{unu1312}) and (\ref{unu1312n}) can be expanded in terms of $x$
and $y$. When $|x|$ and $|y|$ are small, the expansions can be simplified by
ignoring higher order terms. Since in our discussions, parameter $x$ and $y$
are, at most, of order $\mathcal{O}(\lambda_{C})$ where $\lambda_{C}%
=0.2253\pm0.0007$ \cite{pdg} is the Wolfenstein parameter, terms of order
$\mathcal{O}(|x|^{3})$, $\mathcal{O}(|y|^{3})$, or higher can be ignored and,
hence one has
\begin{align}
U_{1} &  \simeq \left(  1-a\right)  K_{1}-y^{\ast}K_{2}-x^{\ast}K_{3}%
,\nonumber \\
U_{2} &  \simeq \left(  1-b\right)  K_{2}+yK_{1}-x^{\ast}yK_{3}%
,\label{unu1312a}\\
U_{3} &  \simeq \left(  1-c\right)  K_{3}+xK_{1},\nonumber
\end{align}
where%
\begin{equation}
a=\left(  |x|^{2}+|y|^{2}\right)  /2,\;b=|y|^{2}/2,\;c=|x|^{2}%
/2.\label{unu1312na}%
\end{equation}

As another example, we consider the case where the perturbation matrix $T$ is
given by a $(1,3)$ rotation followed by a $(2,3)$ rotation. Similarly, one
finds that%
\begin{align}
U_{1}  &  \simeq \left(  1-c\right)  K_{1}-x^{\ast}K_{3},\nonumber \\
U_{2}  &  \simeq \left(  1-b\right)  K_{2}-y^{\ast}K_{3}-xy^{\ast}%
K_{1},\label{unu1323a}\\
U_{3}  &  \simeq \left(  1-a\right)  K_{3}+yK_{2}+xK_{1},\nonumber
\end{align}
where $a$, $b$, and $c$ are given in Eq.~(\ref{unu1312na}).

The perturbation matrix $T$ may depend on two or more different rotations. The
corresponding mixing matrices can be obtained in the same way. This method can
also be used in the case where the perturbation matrix $T$ multiplies $U_{\nu
}^{0}$ from the left, i.e. $U_{\nu}=TU_{\nu}^{0}$ by applying this method to
its transpose, i.e. $\tilde{U}_{\nu}=\tilde{U}_{\nu}^{0}\tilde{T}$.

Obviously, the method discussed above is different from the one dealing with
mixing angles. It can provide a simple way to construct mixing matrices from
LO mixing matrices when the deviations are small. It can also lead to some
interesting results, including the trimaximal-Cabibbo mixing derived in
Sec.~IV. This method can be applied to any LO mixing matrix, but note that
even if the LO mixing matrix $U_{\nu}^{0}$ is in the standard parametrization,
the mixing matrix $U_{\nu}$ may not be in the standard parametrization. This
is not physically significant, although it may require slightly more work to
extract mixing parameters used in the global fits, especially the CP phase. In
addition, in some cases, its physical relevance is more transparent. For
example, when $U_{\nu}$ is given by $TU_{\nu}^{0}$, the perturbation matrix
$T$ can be related to charged lepton corrections.

\section{Tri-bimaximal-Cabibbo mixing}

In the rest of this paper we consider deviations from tri-bimaximal mixing,
i.e.%
\begin{equation}
U_{\nu}^{0}=U_{\mathrm{TBM}}=\frac{1}{\sqrt{6}}\left(
\begin{array}
[c]{ccc}%
2 & \sqrt{2} & 0\\
-1 & \sqrt{2} & -\sqrt{3}\\
-1 & \sqrt{2} & \sqrt{3}%
\end{array}
\right)  .\label{UTBM}%
\end{equation}
It may be instructive to show some simple applications of the method
introduced in the previous section. We begin with the simplest case, where the
perturbation matrix $T$ is described by a single rotation. From
Eq.~(\ref{unu13}), one finds that the second column of $U_{\nu}^{0}$ remains
intact. Hence $U_{\nu}$ is a trimaximal mixing matrix (see, e.g. \cite{TMM}).
Since this case is very simple, below we consider the other two cases
discussed in the previous section.

We consider first the case where $U_{\nu}$ is given by Eqs.~(\ref{unu1323a}).
From the last equation in (\ref{unu1323a}) and Eq.~(\ref{UK}) one has%
\begin{equation}
U_{3}=\left(
\begin{array}
[c]{c}%
\left(  U_{\nu}\right)  _{e3}\\
\left(  U_{\nu}\right)  _{\mu3}\\
\left(  U_{\nu}\right)  _{\tau3}%
\end{array}
\right)  \simeq \left(  1-a\right)  K_{3}+yK_{2}+xK_{1}\label{U31}%
\end{equation}
where $a$ is given in Eq.~(\ref{unu1312na}). Note that, as discussed in the
previous section, in Eq.~(\ref{U31}), terms of order $\mathcal{O}(|x|^{3})$,
$\mathcal{O}(|y|^{3})$ or higher are ignored. Since $U_{\nu}^{0}%
=U_{\mathrm{TBM}}$, then from Eq.~(\ref{UK}) and Eq.~(\ref{UTBM}), one has%
\begin{equation}
K_{1}=\frac{1}{\sqrt{6}}\left(
\begin{array}
[c]{c}%
2\\
-1\\
-1
\end{array}
\right)  ,\quad K_{2}=\frac{1}{\sqrt{3}}\left(
\begin{array}
[c]{c}%
1\\
1\\
1
\end{array}
\right)  ,\quad K_{3}=\frac{1}{\sqrt{2}}\left(
\begin{array}
[c]{c}%
0\\
-1\\
1
\end{array}
\right)  .\label{TMBK}%
\end{equation}
Substituting $K_{i}$ given above into Eq.~(\ref{U31}) leads to%
\begin{equation}
\left \vert y\frac{1}{\sqrt{3}}+x\frac{2}{\sqrt{6}}\right \vert =\left \vert
\left(  U_{\nu}\right)  _{e3}\right \vert =\left \vert \frac{\lambda
e^{-i\varphi}}{\sqrt{2}}\right \vert =\frac{\lambda}{\sqrt{2}}.\label{xytbm20}%
\end{equation}
where $\left(  U_{\nu}\right)  _{e3}$ is parametrized as $\lambda
e^{-i\varphi}/\sqrt{2}$. Parameters $x$ and $y$ can be written as%
\begin{equation}
x=x_{0}\lambda e^{i\varphi}\frac{\sqrt{3}}{2},\quad y=y_{0}\lambda
e^{i\varphi}\sqrt{\frac{3}{2}}.\label{xytbm1}%
\end{equation}
Then one has
\begin{equation}
\left \vert y_{0}+x_{0}\right \vert =1\label{xytbm2}%
\end{equation}
Because $U_{\nu}$ contains only one physical Dirac CP phase, for simplicity we
require that $x_{0}$ and $y_{0}$ are real. In addition, we require that they
do not depend on other parameters. In principal, one may use any $x_{0}$ and
$y_{0}$ as long as $|x|$ and $|y|$ are small. As an example, let $y_{0}=1/3$
and $x_{0}=2/3$. Then from Eqs.~(\ref{unu1323a}) and (\ref{xytbm1}), one has%
\[
U_{\nu}=\left(
\begin{array}
[c]{ccc}%
\sqrt{\frac{2}{3}}\left(  1-\frac{\lambda^{2}}{6}\right)   & \frac{1}{\sqrt
{3}}\left(  1-\frac{5\lambda^{2}}{12}\right)   & \frac{\lambda e^{-i\varphi}%
}{\sqrt{2}}\\
-\frac{1}{\sqrt{6}}\left(  1-\lambda e^{i\varphi}-\frac{\lambda^{2}}%
{6}\right)   & \frac{1}{\sqrt{3}}\left(  1+\frac{\lambda}{2}e^{i\varphi}%
+\frac{\lambda^{2}}{12}\right)   & -\frac{1}{\sqrt{2}}\left(  1-\frac
{\lambda^{2}}{4}\right)  \\
-\frac{1}{\sqrt{6}}\left(  1+\lambda e^{i\varphi}-\frac{\lambda^{2}}%
{6}\right)   & \frac{1}{\sqrt{3}}\left(  1-\frac{\lambda}{2}e^{i\varphi}%
+\frac{\lambda^{2}}{12}\right)   & \frac{1}{\sqrt{2}}\left(  1-\frac
{\lambda^{2}}{4}\right)
\end{array}
\right)  +\mathcal{O}(\lambda^{3})
\]
which is similar to the TBC mixing introduced in \cite{King2}, which is also
parametrized by $\sin \theta_{13}$ whose global-fit value is in good agreement
with the relation $\sin \theta_{13}=\lambda_{C}/\sqrt{2}$. Note that
$\lambda_{C}^{3}\simeq0.011$ and, hence, terms of order $\mathcal{O}%
(\lambda^{3})$ or higher can be neglected. One may check explicitly that
$U_{\nu}$ is unitary up to $\mathcal{O}(\lambda^{3})$ corrections. The mixing
angles extracted from $U_{\nu}$ are given by
\[
\sin^{2}\theta_{12}=\frac{1}{3}-\frac{\lambda^{2}}{9},\; \sin^{2}\theta
_{23}=\frac{1}{2},\; \sin^{2}\theta_{13}=\frac{\lambda^{2}}{2}%
\]
Substituting $\lambda_{C}$ for $\lambda$, one finds that the deviation of
$\sin^{2}\theta_{12}$ from its TBM or TBC value is $\lambda_{C}^{2}/9=0.0056$,
which is much smaller than the experimental error. Note that this is the only
difference between the mixing derived above and the TBC mixing proposed in
\cite{King2}.

The above is an example in which $U_{\nu}=U_{\mathrm{TBM}}T$. Below we
consider another example in which $U_{\nu}=TU_{\mathrm{TBM}}$. Taking the
transpose leads to $\tilde{U}_{\nu}=\tilde{U}_{\mathrm{TBM}}\tilde{T}$.
Therefore, as discussed in Sec.~II, one can use the method in the example
above to obtain $\tilde{U}_{\nu}$, the transpose of $U_{\nu}$. For instance,
one can use Eqs.~(\ref{unu1312a}) to construct $\tilde{U}_{\nu}$. Denote
$\tilde{U}_{\mathrm{TBM}}$ by $(K_{1}^{\prime},K_{2}^{\prime},K_{3}^{\prime})$
and substitute $K_{i}^{\prime}$ for $K_{i}$ in Eqs.~(\ref{unu1312a}), from
$|(\tilde{U}_{\nu})_{3e}|=|(U_{\nu})_{e3}|=\lambda/\sqrt{2}$ one finds that
$|y^{\ast}-x^{\ast}|=\lambda$. As above, one may set $x=x_{0}\lambda
e^{i\varphi}$ and $y=y_{0}\lambda e^{i\varphi}$ and then one has $\left \vert
y_{0}-x_{0}\right \vert =1$. Letting $y_{0}=-x_{0}=1/2$ leads to another
TBC-like mixing matrix with mixing angles given by
\[
\sin^{2}\theta_{12}=\frac{1}{3},\; \sin^{2}\theta_{23}=\frac{1}{2}%
-\frac{\lambda^{2}}{8},\; \sin^{2}\theta_{13}=\frac{\lambda^{2}}{2}.
\]

\section{Trimaximal-Cabibbo mixing}

As we can see, for the TBC-like mixings discussed above, one has $\sin
^{2}\theta_{12}\simeq \sin^{2}\theta_{12}^{\mathrm{TBM}}=1/3$ and $\sin
^{2}\theta_{23}\simeq \sin^{2}\theta_{23}^{\mathrm{TBM}}=1/2$, which fit the
data within the range between the $2\sigma$ and $3\sigma$ experimental bounds.
To achieve a better agreement with the data, corrections to $\theta_{12}$ and
$\theta_{23}$ should also be taken into account. As discussed in the first
section, to avoid introducing additional parameters, we will impose a
phenomenological relation. We note that, to some extent, several well-known
mixing patterns such as TBM and BM can be regarded as phenomenological mixing
patterns. Trimaximal mixing \cite{TMM} and TBC mixing \cite{King2} can also be
derived with certain phenomenological relations in mind. The relation used in
our discussion is given by%
\begin{equation}
\left \vert \left(  U_{\nu}\right)  _{e1}\right \vert ^{2}-\left \vert \left(
U_{\nu}\right)  _{e3}\right \vert ^{2}=\frac{2}{3}\label{u13r}%
\end{equation}
which is consistent with the present data. Like most other phenomenological or
empirical relations, it can hardly be generated as an exact relation.
Nevertheless, note that this relation is also satisfied by TBM and, hence, it
can be considered as a phenomenological constraint on the deviations from TBM
induced by higher order corrections. Its possible theoretical explanation and
phenomenological applications are discussed in the next section.

We discuss first its implication on mixing parameters. As in the previous
section, we denote $\left(  U_{\nu}\right)  _{e3}$ by $\lambda e^{-i\varphi
}/\sqrt{2}$. From Eq.~(\ref{u13r}) we have%
\begin{equation}
\left \vert \left(  U_{\nu}\right)  _{e1}\right \vert ^{2}=\frac{2}{3}%
+\frac{\lambda^{2}}{2},\quad \left \vert \left(  U_{\nu}\right)  _{e2}%
\right \vert ^{2}=\frac{1}{3}-\lambda^{2},\quad \left \vert \left(  U_{\nu
}\right)  _{e3}\right \vert ^{2}=\frac{\lambda^{2}}{2}\label{u13r1}%
\end{equation}
from which it follows that%
\[
\sin^{2}\theta_{13}=\frac{\lambda^{2}}{2},\quad \sin^{2}\theta_{12}\simeq
\frac{1}{3}\left(  1-\frac{5}{2}\lambda^{2}\right)
\]
where higher order terms are neglected. Using the global-fit value of
$\sin \theta_{13}$ given in Eq.~(\ref{gfit}), one finds that the deviation of
$\sin^{2}\theta_{12}$ from $\sin^{2}\theta_{12}^{\mathrm{TBM}}=1/3$ is
$5\sin^{2}\theta_{13}/3=0.038\pm0.0038$, which is much larger than the
$1\sigma$ experimental error. Therefore, to satisfy Eq.~(\ref{u13r}), the
correction to $\theta_{12}^{\mathrm{TBM}}$ cannot be neglected. Also note that
this relation does not constrain the atmospheric mixing angle $\theta_{23}$.
Nevertheless, below we will show that even without making a particular choice
of parameters, the mixing matrix given by Eqs.~(\ref{unu1312a}) leads
straightforwardly to an improved TBC or trimaximal-like mixing that can fit
the data very well, including $\theta_{23}$.

We consider first the case where the perturbation matrix $T$ multiplies
$U_{\mathrm{TBM}}$ from the right, i.e. $U_{\nu}=U_{\mathrm{TBM}}T$. Since
$U_{\nu}^{0}=U_{\mathrm{TBM}}$, from Eq.~(\ref{u13r1}) one finds that
$\left \vert \left(  U_{\nu}\right)  _{e1}\right \vert ^{2}>\left \vert \left(
U_{\nu}^{0}\right)  _{e1}\right \vert ^{2}=2/3$. Because $\left(  U_{\nu}%
^{0}\right)  _{e3}=0$, $\left(  U_{\nu}\right)  _{e1}$ must receive a
contribution from $\left(  U_{\nu}^{0}\right)  _{e2}$, i.e., the first element
of $K_{2}$, which is the second column of $U_{\nu}^{0}$. Therefore we use
Eqs.~(\ref{unu1312a}) to construct $U_{\nu}$, which is the only choice made in
this case. The mixing matrix $U_{\nu}$ can then be derived straightforwardly
in a similar manner as in the example discussed in the previous section. From
the last equation in (\ref{unu1312a}) and $\left \vert \left(  U_{\nu}\right)
_{e3}\right \vert =\lambda/\sqrt{2}$,\ one finds that $|x|\simeq \sqrt{3}%
\lambda/2$. Then from the first equation\ in (\ref{unu1312a}) and
Eq.~(\ref{u13r1}) one has%
\[
\left \vert \left(  1-\frac{|x|^{2}+|y|^{2}}{2}\right)  \sqrt{\frac{2}{3}%
}-\frac{1}{\sqrt{3}}y^{\ast}\right \vert ^{2}\simeq \left(  \frac{2}{3}%
+\frac{\lambda^{2}}{2}\right)
\]
For simplicity, we assume that $y$ is real. Since $|x|\simeq \sqrt{3}\lambda
/2$, it is straightforward to solve the above equation for $y$.
Ignoring\ higher\ order terms,\ one has%
\begin{equation}
x=\frac{\sqrt{3}}{2}\lambda e^{-i\varphi},\quad y=-\frac{3\sqrt{2}}{4}%
\lambda^{2}.\label{tmcpp}%
\end{equation}
Then from Eqs.~(\ref{unu1312a}) it follows that%
\begin{equation}
U_{\nu}\simeq \left(
\begin{array}
[c]{ccc}%
\sqrt{\frac{2}{3}}\left(  1+\frac{3\lambda^{2}}{8}\right)   & \frac{1}%
{\sqrt{3}}\left(  1-\frac{3\lambda^{2}}{2}\right)   & \frac{\lambda
e^{-i\varphi}}{\sqrt{2}}\\
-\frac{1}{\sqrt{6}}\left(  1-\frac{3\lambda e^{i\varphi}}{2}-\frac
{15\lambda^{2}}{8}\right)   & \frac{1}{\sqrt{3}}\left(  1+\frac{3\lambda^{2}%
}{4}\right)   & -\frac{1}{\sqrt{2}}\left(  1+\frac{\lambda e^{-i\varphi}}%
{2}-\frac{3\lambda^{2}}{8}\right)  \\
-\frac{1}{\sqrt{6}}\left(  1+\frac{3\lambda e^{i\varphi}}{2}-\frac
{15\lambda^{2}}{8}\right)   & \frac{1}{\sqrt{3}}\left(  1+\frac{3\lambda^{2}%
}{4}\right)   & \frac{1}{\sqrt{2}}\left(  1-\frac{\lambda e^{-i\varphi}}%
{2}-\frac{3\lambda^{2}}{8}\right)
\end{array}
\right)  +\mathcal{O}(\lambda^{3}).\label{tmc}%
\end{equation}
which is slightly more complicated than the TBC mixing matrix \cite{King2} but
in a better agreement with the data. Hence it can be considered as an improved
TBC mixing. For the reason discussed in Sec. III, terms of order
$\mathcal{O}(\lambda^{3})$ or higher can be neglected. One can also check
explicitly that $U_{\nu}$ is unitary up to $\mathcal{O}(\lambda^{3})$ corrections.

From Eq.~(\ref{tmc}) a simple relation between $\sin \theta_{23}$, $\sin
\theta_{13}$, and the CP phase $\varphi$ can be derived:%
\[
\sin^{2}\theta_{23}\simeq \frac{1}{2}(1+\lambda \cos \varphi+\frac{\lambda^{3}%
}{8}\cos \varphi).
\]
where the $\lambda^{3}$ term can also be neglected. The Jarlskog invariant is
given by
\[
J\simeq-(\lambda \sin \varphi)/6+\mathcal{O}(\lambda^{3}).
\]
Note that $\varphi$ is not the CP phase in the standard parametrization, but
their difference is very small, as will be shown in the next section.

In the case with vanishing CP phase, substituting $\sqrt{2\sin^{2}\theta_{13}%
}$ for $\lambda$ and using the global-fit value for $\sin^{2}\theta_{13}$
given in Eq.~(\ref{gfit}), one has%
\[
\sin^{2}\theta_{12}\simeq0.295_{-0.004}^{+0.004},\quad \sin^{2}\theta
_{23}\simeq0.607_{-0.006}^{+0.005}%
\]
which agree with the data within $1\sigma$ range. Note that by switching
$\lambda \rightarrow-\lambda$ or $\varphi \rightarrow \varphi+\pi$, $\sin
^{2}\theta_{23}$ can be brought into the $1\sigma$ range in the first octant
given in Eq.~(\ref{gfit}). For the nonvanishing phase, one has
\[
0.39\lesssim \sin^{2}\theta_{23}\lesssim0.61
\]
which is within the $3\sigma$ experimental range \cite{Gonzalez}.

In addition, it is interesting to see that $U_{\nu}$ given in Eq.~(\ref{tmc})
may also be regarded as a variant of the $\mathrm{TM}_{2}$ trimaximal mixing
\cite{TMM} with the $\mathrm{TM}_{2}$ condition $|(U_{\nu})_{\alpha2}%
|^{2}=1/3$ ($\alpha=e$, $\mu$, $\tau$) being perturbed by small corrections of
order $\mathcal{O}(\lambda^{2})$. Hence, we refer to it as trimaximal-Cabibbo
mixing since substituting $\lambda_{C}$ for $\lambda$ in Eq.~(\ref{tmc}) leads
to%
\[
\sin^{2}\theta_{12}\simeq0.29,\quad \sin^{2}\theta_{23}\simeq0.61,\quad \sin
^{2}\theta_{13}\simeq0.025
\]
which also agree with the data within $1\sigma$ range.

Now we consider another case where the perturbation matrix $T$ multiplies
$U_{\mathrm{TBM}}$ from the left, i.e. $U_{\nu}=TU_{\mathrm{TBM}}$. As
discussed in previous sections, in this case, one can move the perturbation
matrix to the right by taking transpose. Since the discussion is similar, we
just give the result. From Eqs.~(\ref{unu1312a}) we find that $x$ and $y$
should satisfy%
\begin{align*}
\left \vert -x^{\ast}+y^{\ast}\right \vert ^{2} &  \simeq \lambda^{2},\\
\left \vert 1-\frac{|x|^{2}+|y|^{2}}{2}-x^{\ast}-y^{\ast}\right \vert ^{2} &
\simeq1-3\lambda^{2}.
\end{align*}
One can verify that%
\[
x=-\frac{\lambda e^{-i\varphi}}{2}+\frac{5\lambda^{2}}{8},\quad y=\frac
{\lambda e^{-i\varphi}}{2}+\frac{5\lambda^{2}}{8}%
\]
satisfy the above equations. Then from Eqs.~(\ref{unu1312a}) one has%
\begin{equation}
U_{\nu}\simeq \left(
\begin{array}
[c]{ccc}%
\sqrt{\frac{2}{3}}\left(  1+\frac{3\lambda^{2}}{8}\right)   & \frac{1}%
{\sqrt{3}}\left(  1-\frac{3\lambda^{2}}{2}\right)   & \frac{\lambda
e^{-i\varphi}}{\sqrt{2}}\\
-\frac{1}{\sqrt{6}}\left(  1-\lambda e^{i\varphi}-\frac{9\lambda^{2}}%
{8}\right)   & \frac{1}{\sqrt{3}}\left(  1+\frac{\lambda e^{-i\varphi}}%
{2}+\frac{3\lambda^{2}}{4}\right)   & -\frac{1}{\sqrt{2}}\left(
1-\frac{3\lambda^{2}}{8}\right)  \\
-\frac{1}{\sqrt{6}}\left(  1+\lambda e^{i\varphi}-\frac{11\lambda^{2}}%
{8}\right)   & \frac{1}{\sqrt{3}}\left(  1-\frac{\lambda e^{-i\varphi}}%
{2}+\frac{\lambda^{2}}{2}\right)   & \frac{1}{\sqrt{2}}\left(  1-\frac
{\lambda^{2}}{8}\right)
\end{array}
\right)  +\mathcal{O}(\lambda^{3}).\label{tmc2}%
\end{equation}
which leads to a nearly maximal $\theta_{23}$, i.e.%
\[
\sin^{2}\theta_{23}\simeq \frac{1}{2}(1-\frac{\lambda^{2}}{4}).
\]
Although acceptable, it does not fit the data as well as the
trimaximal-Cabibbo mixing derived above. One may improve that by adjusting $x$
and $y$, which is possible for this case, but the choice of parameters is not
very straightforward, so we will leave that for future considerations when
more experimental data are available.

\section{Summary and Discussions}

In this paper a general method to parametrize the neutrino mixing matrix in
terms of deviations from leading order mixing is discussed. Using this method,
we show that mixing matrices similar to tri-bimaximal-Cabibbo mixing can be
derived by straightforward choices of parameters. However, these mixing
matrices fit the data only marginally. To improve that without increasing the
number of free parameters, we introduce a phenomenological relation, i.e.,
$|\left(  U_{\nu}\right)  _{e1}|^{2}-|\left(  U_{\nu}\right)  _{e3}|^{2}=2/3$.
Two mixing matrices satisfying this relation are constructed. The one referred
to as trimaximal-Cabibbo mixing provides a good two-parameter description of
the present mixing data and, hence, can serve as a useful parametrization as
long as future experimental data do not change the current global fit
significantly. Below we discuss briefly its phenomenological applications and
possible theoretical explanations.

Since the trimaximal-Cabibbo mixing given in Eq.~(\ref{tmc}) involves only two
free parameters, when it is used to parametrize the lepton mixing matrix, the
expressions for many phenomenological quantities can be greatly simplified. As
an interesting application, we consider the neutrino mixing probabilities for
phase-averaged propagation\ with oscillation phase $(\Delta m^{2})L/4E\gg1$.
To simplify our discussion, we use the results given in \cite{TMP} in which
more details can be found. Because in \cite{TMP} the neutrino mixing
probabilities are expressed in terms of a set of parameters different from
those in trimaximal-Cabibbo mixing, one needs to find first the relations
between them. Substituting Eq.~(\ref{tmc}) into the formalism in \cite{TMP}
one finds that the parameters we need can be written as%
\begin{equation}
\epsilon_{21}\simeq-\frac{5\lambda^{2}}{4\sqrt{2}},\quad \epsilon_{32}%
\simeq \frac{\lambda}{2}\cos \varphi,\quad \epsilon_{13}\simeq \frac{\lambda
}{\sqrt{2}}.\label{epij}%
\end{equation}
where $\lambda$ and $\varphi$ are the two parameters in trimaximal-Cabibbo
mixing. In addition, one also needs the Dirac CP phase in the standard
parametrization which is denoted by $\varphi_{D}$. From Eq.~(\ref{tmc}), one
can derive the relation between $\varphi_{D}$ and $\varphi$, which is given
by
\[
\cos \varphi_{D}\simeq \cos \varphi-\frac{1}{2}\lambda^{2}\cos \varphi \sin
^{2}\varphi.
\]
For phenomenological applications, the second term can be ignored since it is
roughly of order $\mathcal{O}(\lambda_{C}^{3})$ and, hence, one may set
$\varphi_{D}=\varphi$. From Eq.~(\ref{epij}) and the results given in
\cite{TMP}, one finds that flavor mixing probabilities can be expressed in
terms of $\epsilon \equiv \lambda \cos \varphi$, i.e.%
\begin{align*}
P_{\nu_{e}\leftrightarrow \nu_{e}} &  =5/9,\quad P_{\nu_{\mu}\leftrightarrow
\nu_{\mu}}=\left(  7+2\epsilon+3\epsilon^{2}\right)  /18,\quad P_{\nu_{\tau
}\leftrightarrow \nu_{\tau}}=\left(  7-2\epsilon+3\epsilon^{2}\right)  /18\\
P_{\nu_{e}\leftrightarrow \nu_{\mu}} &  =\left(  2-\epsilon \right)  /9,\quad
P_{\nu_{e}\leftrightarrow \nu_{\tau}}=\left(  2+\epsilon \right)  /9,\quad
P_{\nu_{\mu}\leftrightarrow \nu_{\tau}}=\left(  7-3\epsilon^{2}\right)  /18.
\end{align*}
The ratio $\Phi_{\mu}/\Phi_{\tau}$ of the $\nu_{\mu}$ flux to the $\nu_{\tau}$
flux arriving at earth, which measures the violation of $\mu-\tau$ symmetry,
can be written as%
\[
\Phi_{\mu}/\Phi_{\tau}=1+26\epsilon^{2}/9.
\]
Comparing with \cite{TMP}, one finds that the expressions for mixing
probabilities are considerably simplified. For more details about mixing
probabilities, see \cite{TMP} and references therein.

Before ending this paper, we briefly discuss possible theoretical explanations
for trimaximal-Cabibbo mixing. As discussed in Sec.~IV, one may begin with a
trimaximal-mixing model. For example, consider the one proposed in \cite{shi}.
It is shown that an $A_{4}$ model with a $1^{\prime}$ (and/or a $1^{\prime
\prime}$) flavon can lead to an effective neutrino mass matrix given by
\begin{align}
M_{\nu}^{0} &  =U_{\mathrm{TBM}}%
\begin{pmatrix}
a+c-\frac{d}{2} & 0 & \frac{\sqrt{3}}{2}d\\
0 & a+3b+c+d & 0\\
\frac{\sqrt{3}}{2}d & 0 & a-c+\frac{d}{2}%
\end{pmatrix}
\tilde{U}_{\mathrm{TBM}}\label{mnu0}\\
&  =a%
\begin{pmatrix}
1 & 0 & 0\\
0 & 1 & 0\\
0 & 0 & 1
\end{pmatrix}
+b%
\begin{pmatrix}
1 & 1 & 1\\
1 & 1 & 1\\
1 & 1 & 1
\end{pmatrix}
+c%
\begin{pmatrix}
1 & 0 & 0\\
0 & 0 & 1\\
0 & 1 & 0
\end{pmatrix}
+d%
\begin{pmatrix}
0 & 0 & 1\\
0 & 1 & 0\\
1 & 0 & 0
\end{pmatrix}
\label{mnu0d}%
\end{align}
where $a$, $b$, $c$, and $d$ depend on model parameters. One can show that
$M_{\nu}^{0}$ leads to $\mathrm{TM}_{2}$ trimaximal mixing with%
\[%
\begin{pmatrix}
|(U_{\nu})_{e2}|\\
|(U_{\nu})_{\mu2}|\\
|(U_{\nu})_{\tau2}|
\end{pmatrix}
=%
\begin{pmatrix}
1/\sqrt{3}\\
1/\sqrt{3}\\
1/\sqrt{3}%
\end{pmatrix}
.
\]
Based on this model, one may introduce higher order corrections or additional
contributions to produce the trimaximal-Cabibbo mixing given by Eq.~(\ref{tmc}).

Before we proceed, one can compare $M_{\nu}^{0}$ with the effective neutrino
mass matrix corresponding to trimaximal-Cabibbo mixing, which is given by%
\begin{equation}
M_{\nu}^{\mathrm{TMC}}\simeq U_{\mathrm{TBM}}%
\begin{pmatrix}
m_{1}+\frac{3}{4}\Delta_{31}\lambda^{2} & -\frac{3\sqrt{2}}{4}\Delta
_{21}\lambda^{2} & \frac{\sqrt{3}}{2}\Delta_{31}\lambda \\
-\frac{3\sqrt{2}}{4}\Delta_{21}\lambda^{2} & m_{2} & 0\\
\frac{\sqrt{3}}{2}\Delta_{31}\lambda & 0 & m_{3}-\frac{3}{4}\Delta_{31}%
\lambda^{2}%
\end{pmatrix}
\tilde{U}_{\mathrm{TBM}}\label{mtmc}%
\end{equation}
where $m_{i}$ are neutrino masses and $\Delta_{ij}\equiv m_{i}-m_{j}$. It can
also be written as%
\begin{equation}
M_{\nu}^{\mathrm{TMC}}=U_{\mathrm{TBM}}%
\begin{pmatrix}
a+c-\frac{d}{2}+\frac{e}{3} & -\frac{\sqrt{2}}{3}e & \frac{\sqrt{3}}{2}d\\
-\frac{\sqrt{2}}{3}e & a+3b+c+d+\frac{2e}{3} & 0\\
\frac{\sqrt{3}}{2}d & 0 & a-c+\frac{d}{2}-e
\end{pmatrix}
\tilde{U}_{\mathrm{TBM}}\label{mtmc0d}%
\end{equation}
where $a$, $b$, $c$, $d$, and $e$ can be determined by comparing the two
equations above. One can show that $M_{\nu}^{\mathrm{TMC}}$ can be decomposed
as%
\begin{equation}
M_{\nu}^{\mathrm{TMC}}=M_{\nu}^{0}+M_{\nu}^{1}=M_{\nu}^{0}+e%
\begin{pmatrix}
0 & 0 & 0\\
0 & 0 & 1\\
0 & 1 & 0
\end{pmatrix}
\label{mtmc0}%
\end{equation}
where $M_{\nu}^{0}$ is the mass matrix given by Eq.~(\ref{mnu0}) or
Eq.~(\ref{mnu0d}). Note that a vanishing CP phase is assumed for simplicity
and the second term in Eq.~(\ref{mtmc0}), i.e., $M_{\nu}^{1}$, can be replaced
by a combination of $\left(
\begin{smallmatrix}
1 & 0 & 0\\
0 & 1 & 0\\
0 & 0 & 1
\end{smallmatrix}
\right)  $, $\left(
\begin{smallmatrix}
1 & 1 & 1\\
1 & 1 & 1\\
1 & 1 & 1
\end{smallmatrix}
\right)  $, $\left(
\begin{smallmatrix}
1 & 0 & 0\\
0 & 0 & 1\\
0 & 1 & 0
\end{smallmatrix}
\right)  $ and $\left(
\begin{smallmatrix}
0 & 0 & 0\\
0 & 0 & 1\\
0 & 1 & 0
\end{smallmatrix}
\right)  $, e.g.~$\left(
\begin{smallmatrix}
0 & 1 & 1\\
1 & 0 & 0\\
1 & 0 & 0
\end{smallmatrix}
\right)  $ or $\left(
\begin{smallmatrix}
1 & 0 & 0\\
0 & 0 & 0\\
0 & 0 & 0
\end{smallmatrix}
\right)  $.

Now one can see that, in the $A_{4}$ model discussed above, if certain
additional contributions can be introduced to account for the difference
between $M_{\nu}^{0}$ and $M_{\nu}^{\mathrm{TMC}}$, trimaximal-Cabibbo mixing
can then be obtained from the latter. For instance, from Eq.~(\ref{mtmc0}) it
follows that $M_{\nu}^{\mathrm{TMC}}$ can be produced by introducing higher
order corrections that contribute dominantly to the $(2,3)$ element of the
effective neutrino mass matrix. On the other hand, by comparing
Eq.~(\ref{mtmc0d}) with Eq.~(\ref{mnu0}), one finds that it can also be
generated by an additional nonvanishing $(1,2)$ mass term in the TBM basis,
which can be obtained by, e.g. adding Higgs triplets (see, e.g. \cite{ishma}).
Nevertheless, in both cases, to suppress other possible contributions, a
certain amount of fine-tuning might be necessary. It would be interesting to
see a more concrete model that can lead to trimaximal-Cabibbo mixing
naturally. We leave that for future work.

\begin{acknowledgments}
This work was supported in part by the National Science Foundation of China
(NSFC) under the grant 10965003.
\end{acknowledgments}

\end{document}